\documentstyle[aps]{revtex}
\begin{document}
\draft
\def\lsim{\lower.5ex\hbox{$\; \buildrel < \over \sim \;$}}
\def\gsim{\lower.5ex\hbox{$\; \buildrel > \over \sim \;$}} 
\title{Gravitational Wave Emission From a Binary Black Hole System in
Presence of an Accretion Disk}
\author{Sandip K. Chakrabarti}
\address{Code 665, NASA Goddard Space Flight Center, Greenbelt, MD, 20771}
\thanks{e-mail: chakraba@tifrvax.tifr.res.in. On leave from Tata Institute
of Fundamental Research, Bombay}
\maketitle

\setcounter{page}{1}
\noindent {Submitted, July 19th, 1995; To appear in Phys. Rev. D.}

\begin{abstract}

We study the  time evolution and gravitational wave emission
properties of a black hole orbiting {\it inside}
an accretion disk surrounding a massive black hole. We  simultaneously
solve the structure equations of the accretion disk in
the presence of heating, cooling, and viscosity as well as the 
equations governing the companion orbit. The
deviation from the Keplerian distribution of the angular momentum of the disk
due to pressure and advection effects causes a significant 
exchange of angular momentum between the disk and the companion.
This significantly affects the gravitational
wave emission properties from the binary system. We show that 
when the companion is light, the effect is extremely important
and must be taken into account while interpreting gravitational
wave signals from such systems.

\end{abstract}


\section*{ I. INTRODUCTION}

The study of gravitational wave emission from binary
systems has received a significant boost in recent years
because of the realization that the detection  of gravitational
waves would directly identify compact and
strongly gravitating bodies, such as neutron stars and black holes.
The Laser Interferometric Gravitational Wave Observatory (LIGO) and 
Laser Interferometer Space Antenna (LISA) project instruments are
being constructed  to achieve these goals \cite{kip87,danz93}.
In order to be able to obtain as accurate information about the
radiating compact bodies as possible, efforts are being made to obtain 
correct forms of quadrupole radiation from a binary system
\cite{poi93,cut93,wis93,blon95}.
In binary systems composed of only neutron stars and stellar black holes,
these computations are adequate. However, when studying effects around
a massive black hole which is assumed to be present in centers
of many galaxies, one needs to consider an additional effect --
the effect of an accretion disk. It is widely believed that 
galactic centers are endowed with massive black holes
\cite{harm94,miyo95},
and in order to explain the observed luminosity from a
galactic core, one needs to supply
matters ranging from  a few hundredths to a few solar masses per year, some of
which may be in the form of stars \cite {rees84}.
Some of the stars could be
compact, namely, neutron stars and stellar mass black holes which
orbit the massive ones at the same time
gradually spiraling in towards the center due to loss of angular momentum
by gravitational waves.

Chakrabarti \cite{chak93} pointed out that the accretion disks close to the
black hole need not be Keplerian and it would affect the
gravitational wave properties. The radiation pressure
dominated disks are likely to be super-Keplerian which would
transfer angular momentum to the orbiting companion and in some
extreme situations, can even stabilize its orbit from coalescing
any further. This was later verified by time-dependent numerical simulations
\cite{mol94}.
When one considers the more general solutions of viscous, transonic, 
accretion disks \cite{chak90,chak96},
one finds that the angular momentum distribution close to the black hole
could be sub-Keplerian as well, depending upon the viscosity 
and the angular momentum at the inner boundary of the disk. 
The disk becomes Keplerian roughly in a distance of
$x_{Kep} \sim (\frac {M^2}{\alpha v})^2$ from the black hole, where, $M=v/a$ is the Mach number
of the flow, $v$ and $a$ are the radial and sound velocities, and 
$\alpha\lsim 1$ is a constant describing the viscosity \cite{ss73}.
Assume that a companion of mass $M_2$ is in an instantaneous
circular Keplerian orbit around a central black hole of mass $M_1$.
This assumption is justified, especially when the orbital radius is
larger than a few Schwarzchild radius where the energy loss per orbit
is very negligible compared to the binding energy of the orbit.
The rate of loss of energy $dE/dt$ in this binary system with an orbital
period $P$ (in hours) is given by \cite{pet63,lang80}
$$
\frac{dE}{dt}=3 \times 10^{33} (\frac {\mu}{M_\odot})^2
(\frac{M_{tot}}{M_\odot})^{4/3} (\frac{P}{1 hr})^{-{10}/{3}} {\rm ergs\ \
sec^{-1}},
\eqno{(1)}
$$
where,
$$
\mu=\frac {M_1 M_2}{M_1+M_2}
$$
and
$$
M_{tot}=M_1+M_2.
$$
The orbital angular momentum loss rate would be,
$$
R_{gw}=\frac{dL}{dt}|_{gw}=\frac{1}{\Omega} \frac{dE}{dt}
\eqno{(2)}
$$
where $\Omega=\sqrt{G M_1/r^3}$ is the Keplerian angular velocity of the
secondary black hole with mean orbiting radius $r$. The subscript `gw'
signifies that the rate is due to gravitational wave emission. 
In presence of an accretion disk co-planer with the 
orbiting companion, matter from the disk [with local
specific angular momentum $l(r)$] will be accreted onto the companion
at a rate close to its Bondi accretion rate \cite{bon52,st84},
$$
{\dot M}_2=\frac{4\pi {\bar \lambda} \rho (GM_2)^2}{(v_{rel}^2+a^2)^{3/2}}
\eqno{(3)}
$$
where $\rho$ is the density of disk matter, ${\bar \lambda}$ is 
a constant of order unity (which we choose it to be $1/2$ for the rest
of the paper), and $v_{rel}=v_{disk}-v_{Kep}$ is the relative
velocity of  matter between the disk and the orbiting companion.
The rate at which the  angular momentum of the companion will be changed 
due to Bondi accretion will be \cite{chak93}, 
$$
R_{disk}=\frac{dL}{dt}|_{disk}={\dot M_2} (l_{Kep} (x) -l_{disk} (x) )
\eqno{(4)}
$$
Here, $l_{Kep}$ and $l_{disk}$ are the local Keplerian and disk
angular momenta respectively. The subscript on the left hand
side signifies that the effect is due to the disk.
If some region of the disk is sub-Keplerian ($l_{disk}<l_{Kep}$), the
effect of the disk would be to reduce the angular momentum of the
companion further and hasten coalescence. If some region of the
disk is super-Keplerian, the companion will gain angular momentum 
due to accretion, and the coalescence is slowed down. In
a {\it thin} disk with a high accretion rate, the Bondi accretion
rate could be very high and the latter effect could, in principle,
stop the coalescence completely \cite{chak93}.

In order to appreciate the effect due to intervention of the
disk, we consider a special case where, $M_2 <<M_1$ and $l_{disk}<<l_{Kep}$. 
In this case,
$\mu \sim M_2$ and $M_{tot}\sim M_1$. The ratio $R$ of these two rates is,
$$
R=\frac{R_{disk}}{R_{gw}}=1.518\times 10^{-7} \frac{\rho_{10}}{{T_{10}}^{3/2}}
{x^4}{M_8}^2
\eqno{(5)}
$$
Here, $x$ is the companion orbit radius
in units of the Schwarzschild radius of the primary,
$M_8$ is in units of $10^8 M_\odot$, $\rho_{10}$ is the density in units
of $10^{-10}$ gm cm$^{-3}$ and $T_{10}$ is the temperature of the
disk in units of $10^{10}$K. It is clear that, for instance, at $x=10$,
and $M_8=10$, the ratio $R\sim 0.015$ suggesting that the effect of the
disk could be a significant correction term to the general relativistic
loss of angular momentum. In the above example, both the disk and the
gravitational wave work in the same direction in reducing the angular
momentum of the secondary. Alternatively,  when $l_{disk} > l_{Kep}$
they act in opposite direction and may slow down the loss of angular 
momentum \cite{chak93}.
In either case, the ratio $R$ is independent
of the mass of the companion black hole, as long as $M_2 <<M_1$.

In what follows, we present equations governing the disk and  the
companion (\S II). In \S III, we solve these equations simultaneously
along with eqs. (2) and (4)
in a few typical cases which show varied nature of the disk
structure and evolution of the companion. These disks are the
generalization of the viscous, isothermal disks
obtained earlier \cite{chak90}.
We also present some interesting
observations on non-axisymmetric disks containing spiral shocks. Finally, in
\S IV, we make concluding remarks.

\section*{II. GOVERNING EQUATIONS}

To simplify the equations governing the accretion disks, we make the
usual assumption that the disk is thin, so that the vertical averaging 
of density, pressure, and viscous stress, could be
done and the vertical velocity component can be ignored.  We assume the
disk to be axisymmetric, an assumption relaxed in studying
non-axisymmetric disks. Instead
of solving fully general relativistic equations, we assume the
Paczy\'nski-Wiita \cite{pw80} potential approach which is sufficiently accurate
to describe physical quantities around a Schwarzschild black hole,
provided one is not too close to the horizon ($r \gsim r_g=2GM_1/c^2$
is the Schwarzschild radius of the primary black hole). In this
approach, the potential of the central body $\Phi(r)=-GM_1/r$
is replaced by a pseudo-Newtonian potential $\Phi_{PN} (r)=-GM_1/(r-r_g)$.
The effect we are presently discussing, namely the transport of angular momentum
from the disk to the companion, is not of general-relativistic origin.
Hence the major conclusions should not be affected by our simplified
approach. 

The steady state accretion disk equations are:

\noindent (a) The radial momentum equation:

$$
v \frac{dv}{dx} +\frac{1}{\rho}\frac{dP}{d\rho} 
+\frac {l_{Kep}^2-l^2}{x^3}=0,
\eqno{(6a)}
$$

\noindent (b) The continuity equation:

$$
\frac{d}{dx} ( \rho x h v) =0 ,
\eqno{(6b)}
$$

\noindent (c) The azimuthal momentum equation:

$$
v\frac{d\l(x)}{dx} -\frac {1}{\rho x h}\frac{d}{dx} 
(\frac{\alpha P x^3 h} {\Omega_{Kep}} \frac{d\Omega}{dx}) =0
\eqno{(6c)}
$$

\noindent (d) The entropy equation:

$$
\Sigma v T \frac{ds}{dx} = Q^+ - Q^- .
\eqno{(6d)}
$$

Here $l_{Kep}$ and $\Omega_{Kep}$ are the Keplerian angular momentum and
Keplerian angular velocity, respectively, $\Sigma$ is the  density $\rho$
vertically integrated, $h=h(x)$ is the half-thickness of the disk at radial
distance $x$, $v$ is the radial velocity, $s$ is the entropy density
of the flow and $Q^+$ and $Q^-$ are the heat gained and lost by the flow. 
We compute $h(x)$ assuming the disk is in a hydrostatic balance equation
in thethe  vertical direction.
$l(x)$ is the angular momentum distribution of the disk matter.
Here, we have chosen geometric units, thus $x=r/r_g$ is distance
in units of a Schwarzschild radius, $l(x)$ is in units of $2GM_1/c$,
and velocities are in units of the velocity of light. We have implicitly
assumed $M_2 <<M_1$ so that the gravitational effects due to the 
companion in shaping the disk could be ignored. However, locally,
the companion is capable of exerting its effect to accrete matter from the
disk. $\alpha\lsim 1$ in the above equation is the viscosity parameter of 
Shakura  and Sunyaev \cite{ss73}, which is widely used to describe the viscous 
stress: $w_{r\phi}=-\alpha P$. This stress transports angular momentum from
the inner to the outer region of the disk. We choose total pressure
(thermal plus ram) in this prescription in order that the angular
momentum remains continuous across shock waves as well \cite{chak96}.

The equation governing the companion, treated as a test particle
in the field of the massive black hole is simply
$$ 
(\frac{dx}{dt})^2 = - \frac {1}{x-1} + \frac{l_{Kep}^2}{x^2} .
\eqno{(7)}
$$

In the following section, we present simultaneous solutions of the
four sets of equations: (2), (4), 6(a-d) and (7)
obtained by a very accurate fourth-order Runge-Kutta method \cite{numres}.

We note from the estimate of the ratio $R$ in (5) that it is independent
of the mass of the companion. However, the evolution time scale
$L/(dL/dt)$ of the companion orbit depends inversely upon its mass $M_2$.
Thus, even when the effect is very small $R<<1$, the smaller
companion will evolve so slowly that the number of cycles will be
significantly affected. 
While integrating the above equations, however, we had to consider the computing
ability of our machines. This constrained us to study 
cases with faster evolution time scale: we chose $M_1=10^8 M_\odot$
and $M_2=10^6M_\odot$. Second, we note from (5) that the effect is
directly proportional to the density of the gas in the disk, which in 
turn depends upon the accretion rate ${\dot M}_1$ of the primary.
It is customary to express accretion rates in astrophysics in units
of Eddington rates ${\dot M}_{Edd}= 4\pi GM_1 m_p/\sigma_T\sim 0.2 M_\odot M_8$
yr$^{-1}$, where $m_p$ is the proton mass and $\sigma_T$ is the Thomson
cross section. Unlike the accretion process onto ordinary
stars radiating from its surface, a black hole accretion process
need not be limited by its Eddington rate. Since at the most $\eta=0.06$
fraction of the rest mass energy is released by accreting matter on a 
Schwarzschild black hole \cite{st84},
a critical 
rate of ${\dot M_{crit}}\sim {\dot M_{Edd}}/{\eta}\sim 16 {\dot M_{Edd}}$
is very reasonable to choose. For concreteness, we assume ${\dot M}_1 \sim
70 {\dot M}_{crit}$ (i.e., ${\dot M}_1 \sim 1000{\dot M_{Edd}}$). The effects
we describe will be proportionately weaker when smaller rates are used.

To keep the problem simple enough, we have considered
the companion orbits to be circular. The general elliptic orbit is easily
studied by including the evolution of the azimuthal coordinate
in conjunction with eq. (7). This will be done later. Second,
we assume that the companion orbit is co-planer with the disk, so that
the companion is always immersed inside the disk. When it is not so,
one has to include the fraction of time the companion is exchanging
angular momentum with the disk and the effect would be proportionately
reduced as well. In the case of lighter companions, the time scale of evolution
is very long, and it is not unlikely to imagine that the orbits originally
away from the equatorial plane will gradually lose the momentum component
normal to the disk by repeated interaction \cite{syer91} and eventually
come to the plane of the disk much before our effects become important.
In the case of massive companions, the evolution due to gravitational waves
could be very rapid, and they may remain off the plane, as is
possibly the case for OJ287 \cite{silan88}. 

\section*{III. Simultaneous Solutions of the Governing Equations}

Before we present the results of our investigation, we
discuss briefly what type of accretion disk solutions are expected.
From (6c) (also see Chakrabarti \cite{chak90})
one observes that a weakly viscous disk starts deviating
from Keplerian distribution very far away, whereas the strongly viscous disk
remains Keplerian until close to the black hole. The disk cannot
remain Keplerian very close to the black hole, as the velocity
increases and the advection term [first term in (6a)] becomes 
important. Similarly, when the accretion rate is high or very low,
the radiation pressure \cite{pw80} or the gas pressure \cite{reesnat} 
becomes important and the pressure term [second term in (6b)]
cannot be ignored. Both of these terms were ignored in the study
of Keplerian disks \cite{ss73,nt73}.
Also, in a Keplerian disk, angular momentum $l(r)=l_{Kep}$ is used 
independent of viscosity prescription. But generally, for some ranges 
of viscosity and accretion rates, this need not be true. Since our effects 
are non-zero only in non-Keplerian disks (eq. 4), it is essential
that we include these effects. Near $x=x_{Kep}$ where $l(x)\sim l_{Kep}$,
the distribution rapidly deviates from Keplerian to highly sub-Keplerian.
This causes a `micro-burst' of the gravitational wave emission as we
shall show below.

An important class of stable solutions of 6(a-d) involve shock waves 
\cite{chak90,chak96}
where the centrifugal barrier of the flow brakes the radial motion of the disk
before the disk can continue through the sonic point to become
supersonic, thus satisfying the boundary condition on the
horizon. At the shock wave, the density, velocity and temperature
change discontinously and the effect we are considering is expected to be 
discontinuous as well. This causes a `micro-glitch' in the gravitational wave. 
In the case of non-axisymmetric disks, the spiral shocks cause
micro-glitches to appear repeatedly depending on the number of 
spiral arms. This will be discussed towards the end of this Section.

A few cases of the solutions are presented here, which 
cover all possible types of solutions. All the disk solutions are
characterized by three parameters (instead of four, since the flow
passes through one or more sonic points \cite{chak90}). These
parameters are $l_{in}$ [angular momentum at the inner edge
of the disk; the integration constant of equation (6c)],  ${\dot M}_1$ 
(accretion rate on the primary; the integration constant of eq. 6b) 
and $x_{in}$ (the location of the inner sonic point; this
defines specific energy of the flow at a given point). Alternatively, we 
could choose $x_{Kep}$, the location where the disk starts deviating from the
Keplerian distribution, but we prefer to choose the sonic point
location for convenience. We also choose an $\alpha$, the
unknown viscosity parameter, the polytropic index of the gas $\gamma$
which defines the specific internal energy of an ideal gas: $e=(\gamma-1)^{-1}
P/\rho$ and the $Q_+ - Q_-$, the relative importance of cooling
and heating [eq. (6d)]. These are not completely independent parameters, 
but to obtain them one requires to include other equations in the list
6(a-d) to describe the viscous mechanism (such as 
poorly understood turbulence and convections), and
cooling processes (such as the Compton effect, bremsstrahlung, pair
creations and annihilations etc.). Instead of bringing 
in these equations we choose reasonable values for these quantities.
Since the ratio $R$ [eq. (5)] does not depend on $M_2$
(but the orbital evolution time does)
we consider only the case of $M_1=10^8M_\odot$
and $M_2=10^6 M_\odot$ in order to hasten the evolution of the companion orbit. 

\noindent Case A: Figures 1(a)-c) show results 
where the disk always remains sub-Keplerian after deviating
from the Keplerian disk at $x_{Kep}\sim 90 r_g$. The disk smoothly
passes through the inner sonic point at $x_{in}=2.3$. Other parameters
are $l_{in}=1.7$, ${\dot M}_1=1000{\dot M}_{Edd}$, $\gamma=5/3$,
$\alpha=0.02$, $Q_-=Q_+$. In Figure 1a, we notice that the flow
quickly becomes highly sub-Keplerian first. However, before
entering the black hole it becomes only moderately sub-Keplerian.
Figure 1b shows the ratio $R=R_{disk}/R_{gw}$. The ratio $R$
jumps to almost $0.1$ around $x=80$ before decreasing to a very
small value close to a black hole. Figure 1c shows the number of times
the companion orbits the primary (twice the number of full
gravitational waves emitted). The solid curve is drawn including the effect
of the accretion disk, while the dashed curve is drawn considering the
usual binary orbit evolution (eq. 2) without the presence of the
disk. Time in the abscissa denotes time passed since the companion entered
the sub-Keplerian disk. Two effects are clear: (a) the binary coalescence
takes place roughly $10\%$ times faster and (b) the number of 
orbital cycles is also about $10\%$ times higher at the time of 
coalescence. If an accretion rate of ${\dot M}_1={\dot M}_{Edd}$ 
were chosen instead,
the effect would be reduced by a factor of $1000$. If a lighter
black hole of $M_2\sim M_\odot$ was chosen instead, a longer
orbital evolution time due to weaker gravitational wave loss gives 
rise to the same effect.

\noindent Case B: In this case we choose a solution with a standing
shock wave. 
The disk parameters are: $l_{in}=1.6$, $x_{in}=2.87$, ${\dot M}_1=1000{\dot
M}_{Edd}$, $\gamma=4/3$, $\alpha=0.05$ and $Q_-=0.5 Q_+$.
Figure 2a shows the Mach number variation as a function of distance from the
black hole. The arrowed curve is followed by the disk after it deviates
from Keplerian disk at $x_{Kep}=480 r_g$. The disk first passes through the
outer sonic point (located at $x_{out}=50$), then through the shock at 
$x_s=13.9$
and finally enters the black hole through the inner sonic point at $x_{in}=
2.87$. The shock location or the location of the outer
sonic point is not a free-parameter, but are self-consistently
determined from the Rankine-Hugoniot relation \cite{chak89,chak90}.
The shock solution is always chosen if it is available to the 
disk, since the entropy at the inner sonic point is higher compared
to its value at the outer sonic point, and the required entropy
must be generated at the shock. Figure 2b shows the ratio of disk
to Keplerian angular momentum distributions. Figure 2c shows the ratio
$R$ as a function of the distance. The ratio becomes almost $5$,
($5 \times 10^{-3}$ for ${\dot M}_1={\dot M}_{Edd}$)
a micro-burst of a sort, around $x=400r_g$. There is also a 
glitch at the shock location. In cases with a stronger
shock wave the glitch would be stronger. 

\noindent Case C: In this case we choose disk parameters so as to 
obtain a super-Keplerian region in the disk. We choose $l_{in}=1.88$,
$x_c=2.2$, ${\dot M}_1=1000{\dot M}_{Edd}$, $\alpha=0.005$, $\gamma=4/3$
and $Q_-=Q_+$. The disk deviated from Keplerian disk at 
$x_{Kep}=7.5r_g$. Figure 3a shows the ratio of disk to Keplerian distributions
which clearly shows the sub-Keplerian as well as super-Keplerian regions.
Figure 3b shows the ratio $R$  (eq. 5).
The fractional change in orbital cycle number with and without the disk is
$\frac{\delta N}{N}\sim R$. Thus, $\delta 
N \sim 1$ only when $N \sim R^{-1} \sim \frac{1}{p}
\frac{l(x_{Kep})}{{dl/dt}|_{gw}}$. Thus, even if $R$ is small,
lighter companions should survive long enough to feel the effect
of angular momentum exchange. Chakrabarti \cite{chak93} considered a 
thin disk where the density of the disk was high enough to stabilize the
companion orbit in the super-Keplerian region.

\noindent Case D: In this case we solve non-axisymmetric disk equations
\cite{spirchak} instead of eqs. 6(a-d). Here spiral shocks formed
would produce repeated glitches in the gravitational waves pattern. 
The simplest solutions of the non-axisymmetric shocks are obtained
by assuming self-similarity in $x$ and all the disk velocity components
vary as $q_i(\phi) x^{-1/2}$ and the density of the disk varies as
$q_\rho (\phi) x^{-3/2}$. Here, azimuthally varying coefficients
$q_i (\phi)$ and $q_\rho (\phi)$ are to be determined
from boundary conditions. Figure 4a shows a typical solution for the
velocity coefficients when a two armed spiral shock solution is 
considered. The solid, long-dashed and short-dashed curves show the
the radial, azimuthal and sound velocity coefficients and the 
dotted curves show the density coefficient as they vary with the
azimuthal angle. The shocks are located at $\phi=0$ and $\phi=\pi$.
In a Keplerian disk, the azimuthal velocity coefficient would be unity
throughout the disk. In this example, the azimuthal velocity
coefficient varies from $92\%$ of Keplerian to $26\%$ 
Keplerian as the flow crosses the shock front. Other components
also suffer a jump. In a single circular orbit, the companion thus
passes twice through these jumps. Figure 4b shows
(in arbitrary units) the glitches in the ratio $R$ in a single orbit.
In an axisymmetric disk,
the glitch appears only once, but in a disk with spiral shocks the effect
occurs repeatedly and cumulative effect becomes important due to
the repeated passage of the companion through the shock.
This could significantly  modify the shape of the gravity wave signals.

\section*{ IV. CONCLUDING REMARKS}

It is widely recognized that accurate templates
of possible signals may be essential to determine the nature of 
radiating compact bodies \cite{kip87}. 
In this paper, we have discussed several important ways a gravitational
wave signal from a binary companion could be modified in presence of an
accretion disk. We find that even under very normal circumstances,
the effects will be sufficiently significant and our effect 
may influence the templates constructed assuming the
absence of accretion disks. 

In a binary system containing lighter mass black hole components, the 
accretion disk need not be present. Systems involving
massive black holes at the galactic center may necessarily
contain accretion disks and lighter black hole companions.
The frequency of the gravitational wave $f_{gw}=2.25\times 10^{-4} 
x^{-3/2} M_8^{-1}$ is well outside LIGO sensitivity, but could
be well within LISA sensitivity \cite{danz93,hils95}.
By self-consistently solving the equations governing the accretion disk 
structure and the evolution of the binary orbit, we first showed that
accretion disks close to the black hole are in general {\it non-Keplerian}.
In the sub-Keplerian region of the disk, the residence time
of the companion inside a disk and the probability of its observation
would be reduced. On the other hand, the super-Keplerian region
enhances the residence time and the probability of observing these
systems is higher. We also find that the orbital evolution may be faster
away from the black holes where the disk angular momentum distribution
starts deviating from Keplerian distribution. The population density of compact
stars close to galactic nuclei should be affected by their
interaction with the disk. These effects should be taken into
consideration while determining the band of maximum sensitivity
of future instruments for gravitational wave astronomy.

The discussions made in this paper involving a companion black hole
are valid even when a neutron star is chosen instead and may approximately
remain valid when an ordinary star orbits the central black hole. In the latter
case, the angular velocity of the disk changes significantly along a
radial direction across the star. This would cause
some angular momentum of the disk to spin up or spin down the star itself rather
than changing its orbital angular momentum. Furthermore,
the star may lose angular momentum through winds. Therefore, our result need not be 
strictly valid in these systems. These effects are negligible if the companion 
is a black hole or a neutron star because of its small size and the absence of winds. 

The assumption of a thin disk in vertical equilibrium 
(namely, that the vertical velocity is negligle compared to the radial or azimuthal velocity)
enabled us to integrate the governing equations.
Numerical simulations of fully three dimensional disks \cite{mlc94} indicate that
the assumption of the vertical equilibrium adequately describe the disk properties and therefore
we do not believe that the conclusions drawn in the present paper are
affected by this 
assumption. Another implicit assumption has been that the disk remains continuous
(and does not break apart in the form of rings as in the case of orbiting matter around Saturn)
even in the presence of an orbiting companion. The formation of gaps in the
disk is possible only if the instantaneous gap is not filled in by
the accreting matter through radial pressure or viscous forces 
\cite{syer91,linpa}.
This implies that either the Roche radius ($R_L$) of the star orbiting at radius $r$ is 
greater than the disk thickness,
$R_L \sim (M_2/M_1)^{1/3} r \gsim h \sim a r^{3/2} \sim 0.5 r$ (here, the
sound speed
$a \sim 1/\sqrt{3} r^{-1/2}$), or, the viscosity parameter is so small that the angular
momentum transfer rate by tidal coupling through the satellite is higher than that
by viscosity: $\alpha <1/40 (M_2/M_1)^2 (r/h)^5 \sim 3/8 (M_2/M_1)^2$. It is clear that
for the cases we are interested in, namely, for $M_2/M_1 \lsim 10^{-6}$, and $\alpha \gsim 10^{-3}$,
neither of these conditions would be satisfied. Thus, we do not think that gaps would be
formed by orbiting black holes or neutron stars. 

Our goal in this paper has been to indicate a new physical effect which
may change the gravitational wave pattern significantly.
Construction of accurate templates for inferring component masses 
of the gravitating systems is beyond the scope of the present paper. 
It is possible that one could estimate the mass of the central black hole by 
comparing the observed optical and UV radiation spectra 
and the hard and soft X-rays with the
theoretically derived spectra using these general disk models \cite{chak96,chaktitu}. 
The accretion rate of the Keplerian disk could be inferred from the normalization of the optical and UV flux
as well. Thus two parameters are easily eliminated. Viscosity is not an well understood 
process in the context of accretion phenomena, but typical
values of the parameters have been presented in the literature from time to time \cite{visni,balbu}
which we have considered here for simplicity. Since whether the flow is Keplerian or non-Keplerian
depends very crucially upon the viscosity parameter \cite{chak96}, undoubtedly, 
a complete resolution of the present problem is hinges upon a better understanding of the 
viscosity of the accretion disk.

\section*{ACKNOWLEDGMENTS}

The author is thankful to Kip Thorne and C. Cutler for helpful comments.
The research has been supported by National Research Council
through a senior research associateship award.

\newpage

\section*{FIGURE CAPTIONS}

\begin{figure}
\noindent FIG. 1a: Ratio of disk angular momentum distribution
to Keplerian distribution of a disk which is entirely
sub-Keplerian for $x<x_{Kep}=90$. See text for flow parameters.
\end{figure}

\begin{figure}
\noindent FIG. 1b: Ratio of the rates of change of angular momentum
of the companion due to exchange with the disk and due to
gravitational wave emission. The ratio is highest 
in regions closer to the Keplerian boundary.
\end{figure}

\begin{figure}
\noindent FIG. 1c: Comparison of number of orbital cycles in a binary
with disk (solid) and without disk (dashed) as time passes
since the companion enters the sub-Keplerian region of the disk.
Companion falls faster when the disk is present.
\end{figure}

\begin{figure}
\noindent FIG. 2a: Variation of Mach number of the disk which includes
a shock wave at $x_s\sim 13.9$. The arrowed curves are the 
solutions chosen by the flow. See text for flow parameters.
\end{figure}

\begin{figure}
\noindent FIG. 2b: Ratio of disk angular momentum distribution
to Keplerian distribution of the disk with a shock
which is entirely sub-Keplerian for $x<x_{Kep}=480$.
\end{figure}

\begin{figure}
\noindent FIG. 2c: Ratio of the rates of change of angular momentum
of the companion due to exchange with the disk and due to
gravitational wave emission. The ratio is highest 
in regions closer to the Keplerian boundary. Note the glitch
at the shock location which could be very high for stronger shocks.
\end{figure}

\begin{figure}
\noindent FIG. 3a: Ratio of disk angular momentum distribution
to Keplerian distribution of the disk where
the sub-Keplerian disk below $x<x_{Kep}=7.5$ becomes super-Keplerian
close to the black hole. See text for flow parameters.
\end{figure}

\begin{figure}
\noindent FIG. 3b: Ratio of the rates of change of angular momentum
of the companion due to exchange with the disk and due to
gravitational wave emission. The ratio is highest 
in regions closer to the Keplerian boundary. Note the 
change in sign of the ratio as the companion enters the
super-Keplerian region.
\end{figure}

\begin{figure}
\noindent FIG. 4a: Velocity and density 
variations with azimuthal angle in a non-axisymmetric
disk with a two-armed spiral shock waves located at $\phi=0$ and $\phi=
\pi$. The velocities are in units of local Keplerian velocity, while
the density is in an arbitrary unit.
\end{figure}

\begin{figure}
\noindent FIG. 4b:Ratio of the rates of change of angular momentum
of the companion due to exchange with the disk and due to
gravitational wave emission. The jump in the ratio at the spiral shocks
produces glitches twice per orbital cycle (once per gravitational
wave signal).
\end{figure}

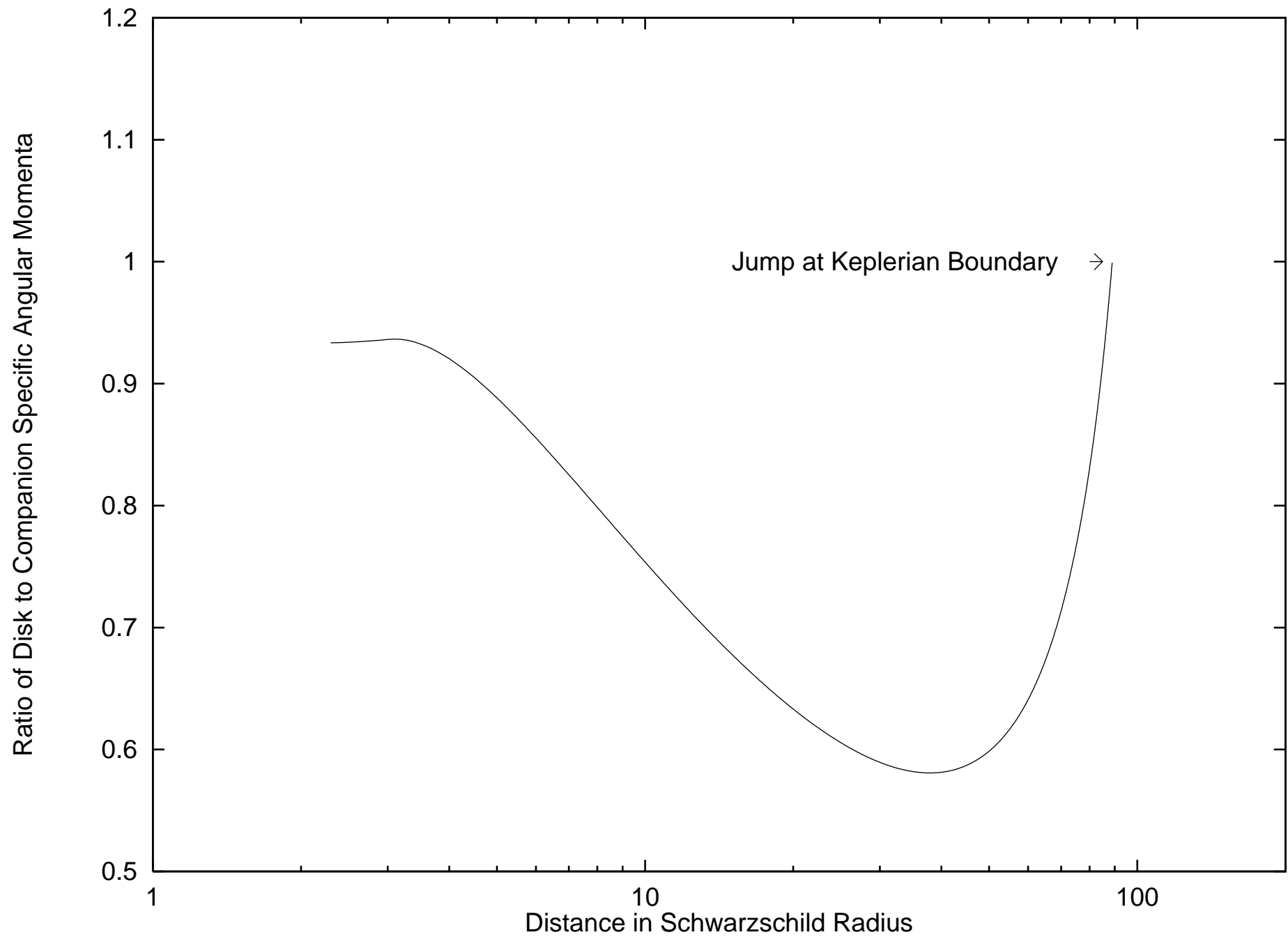
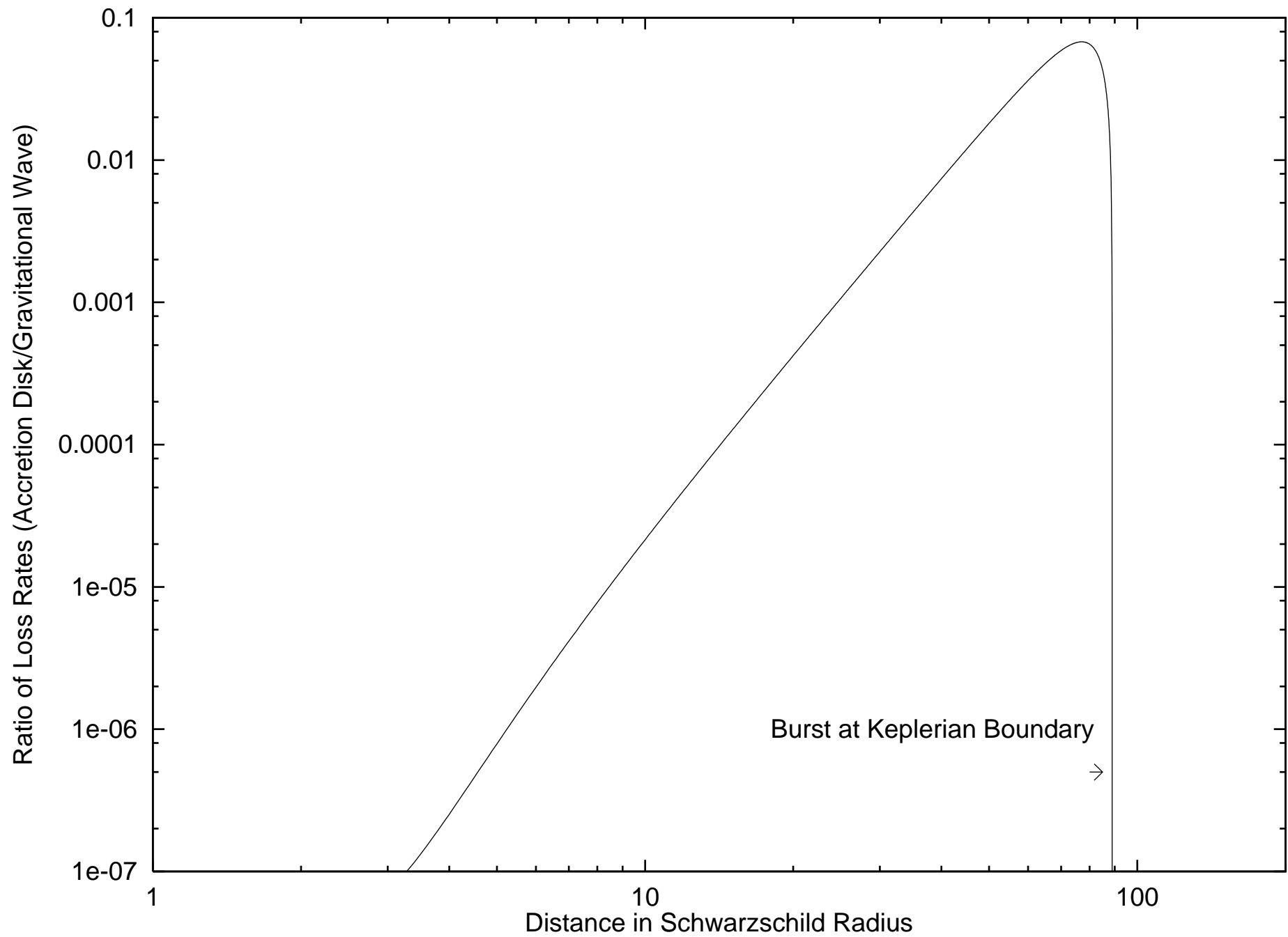
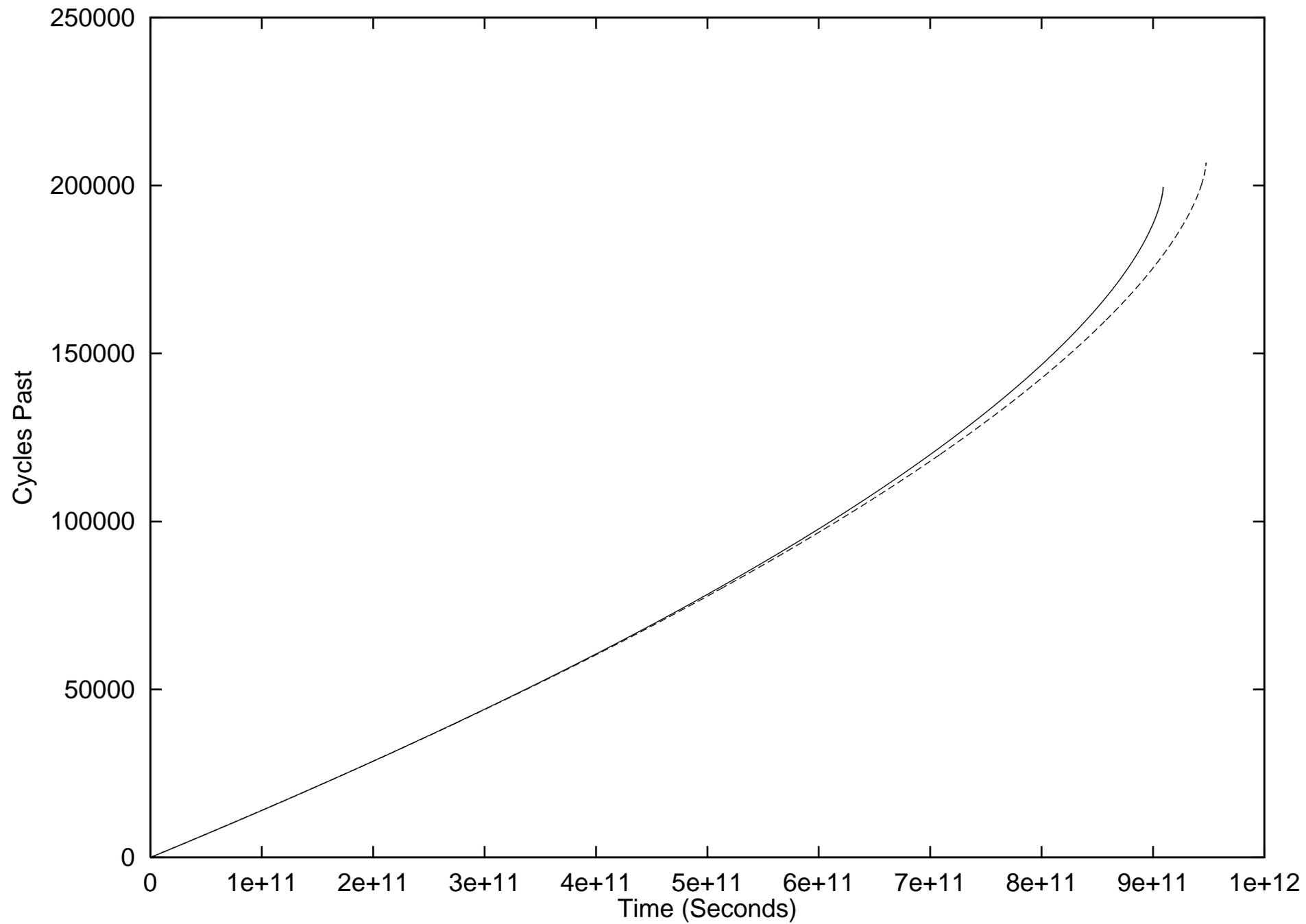
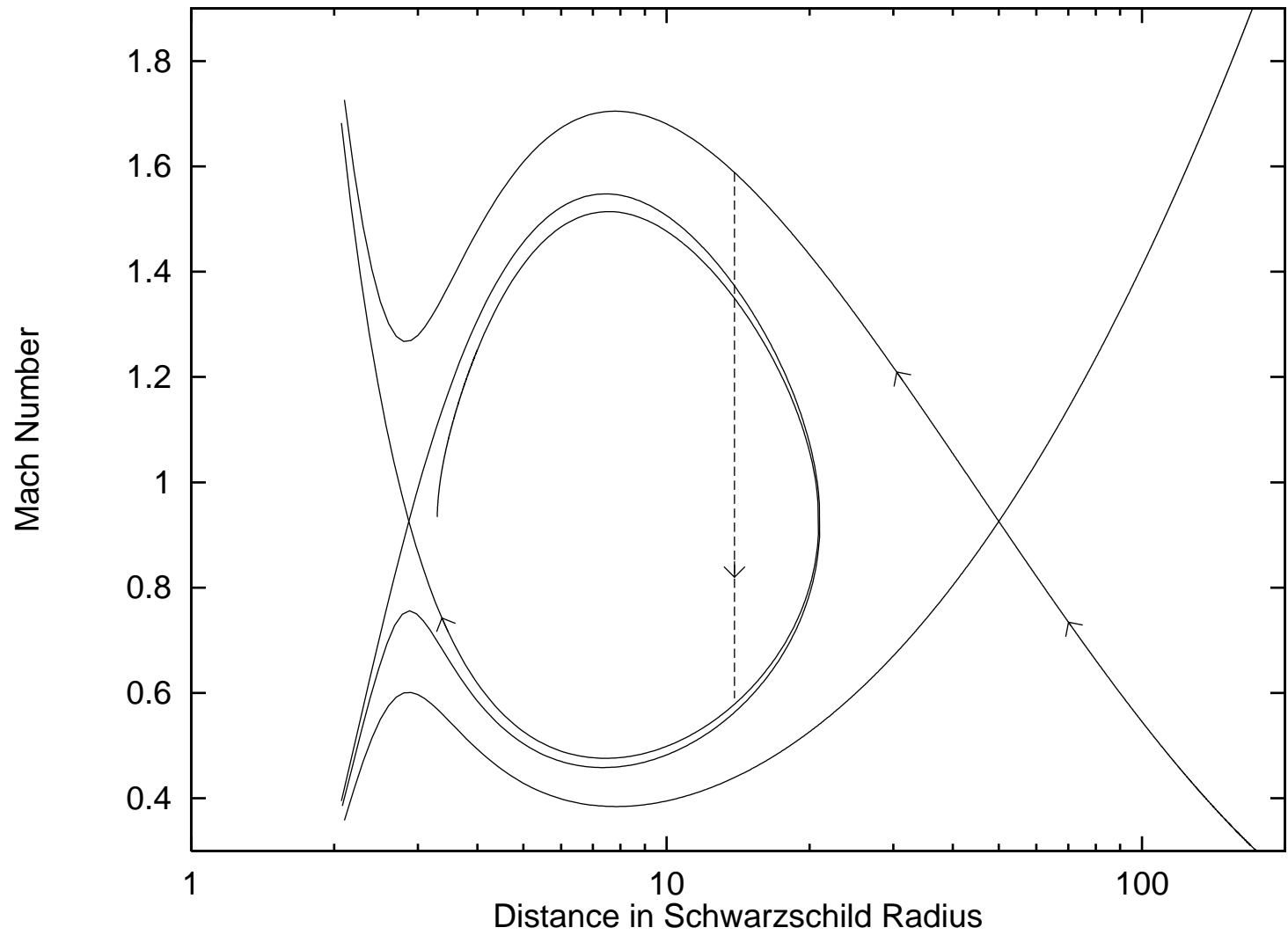
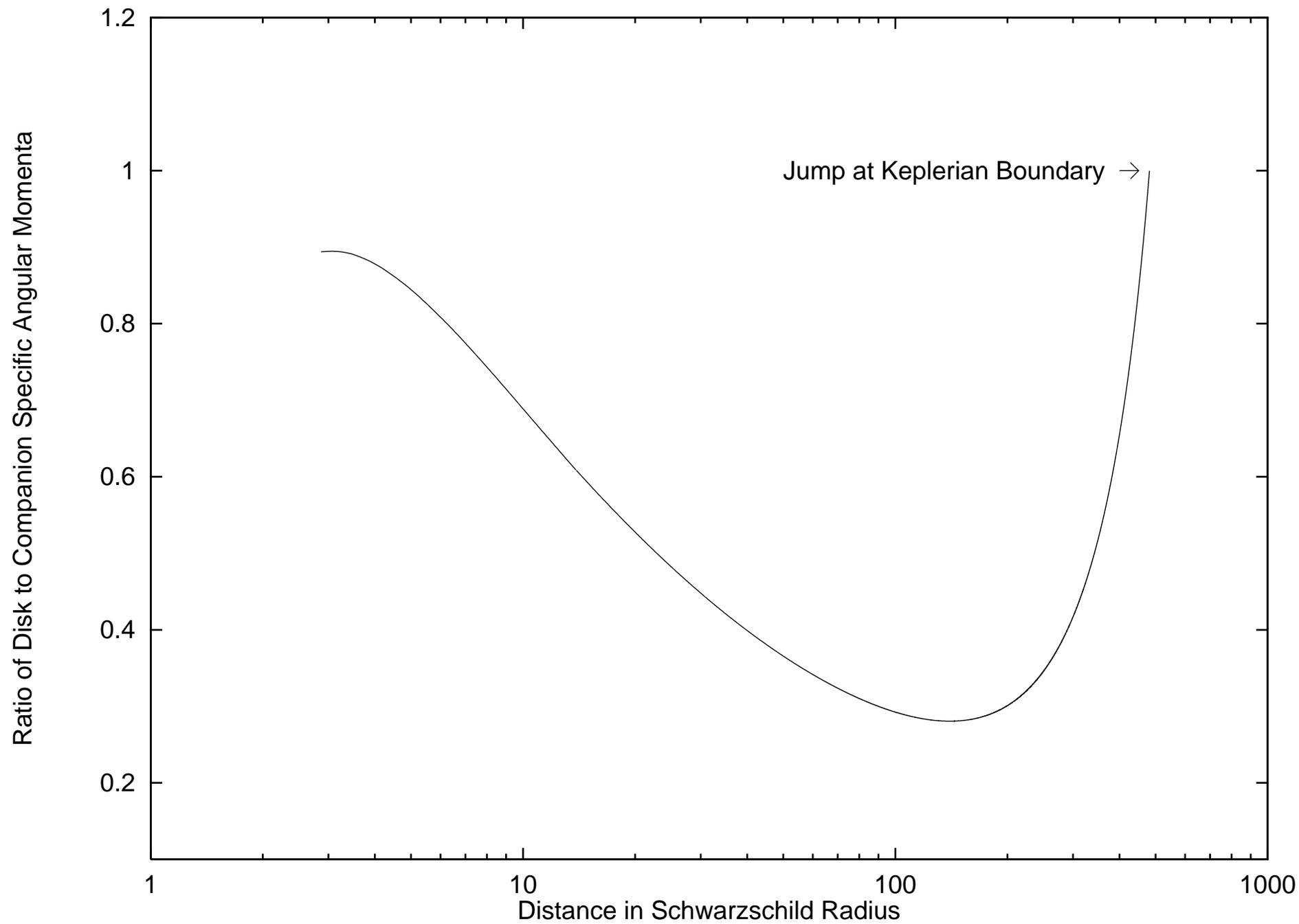
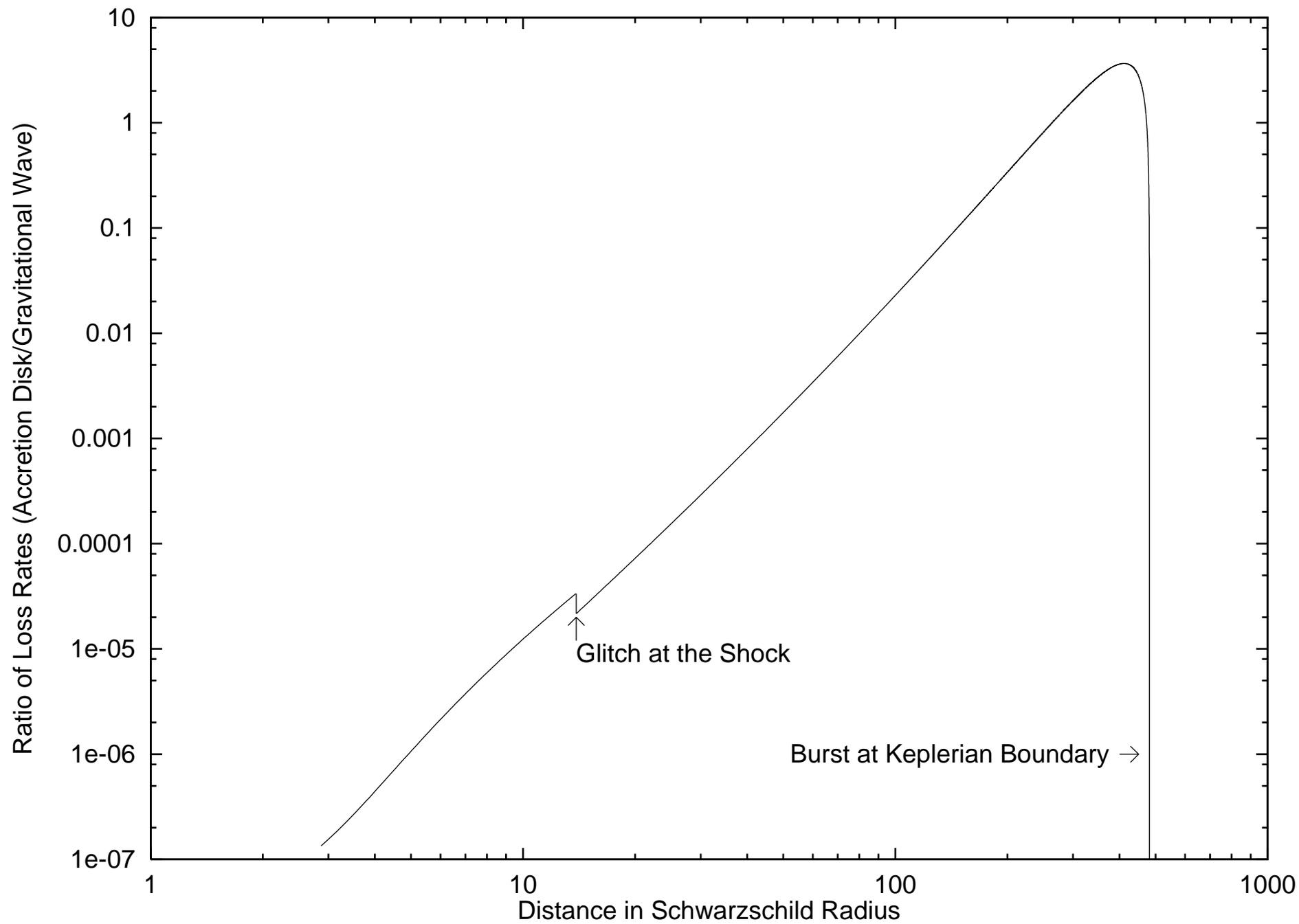
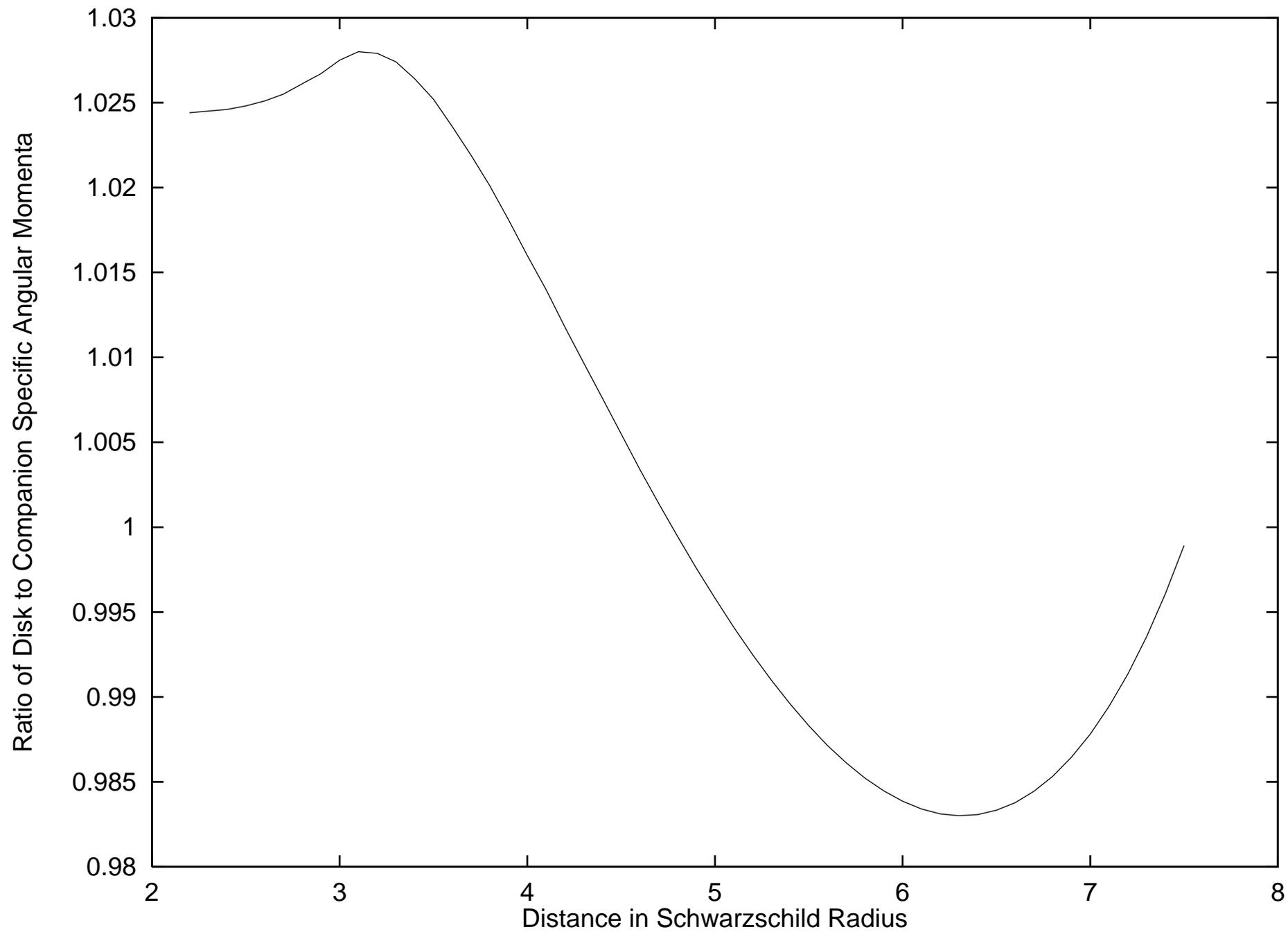
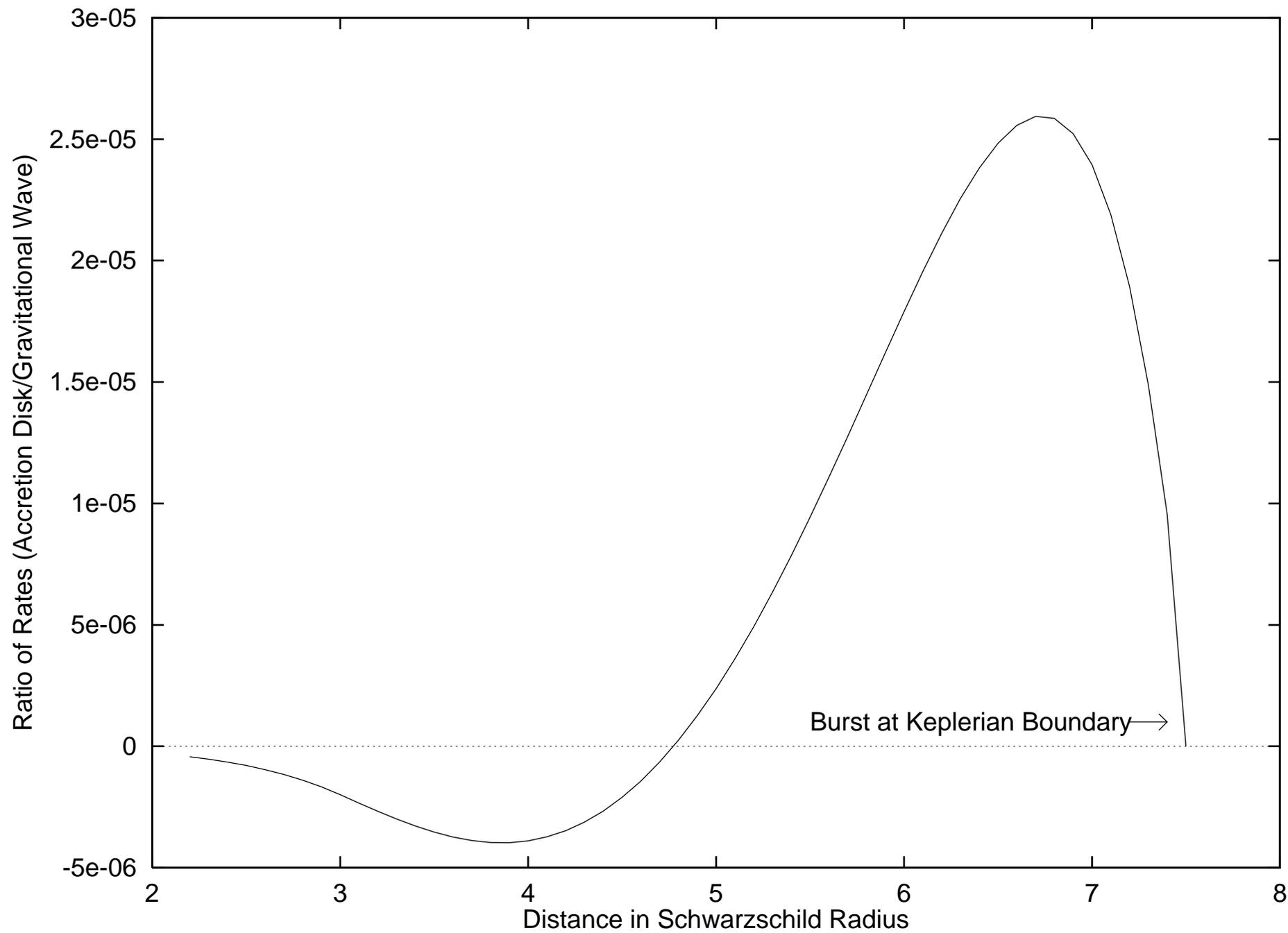
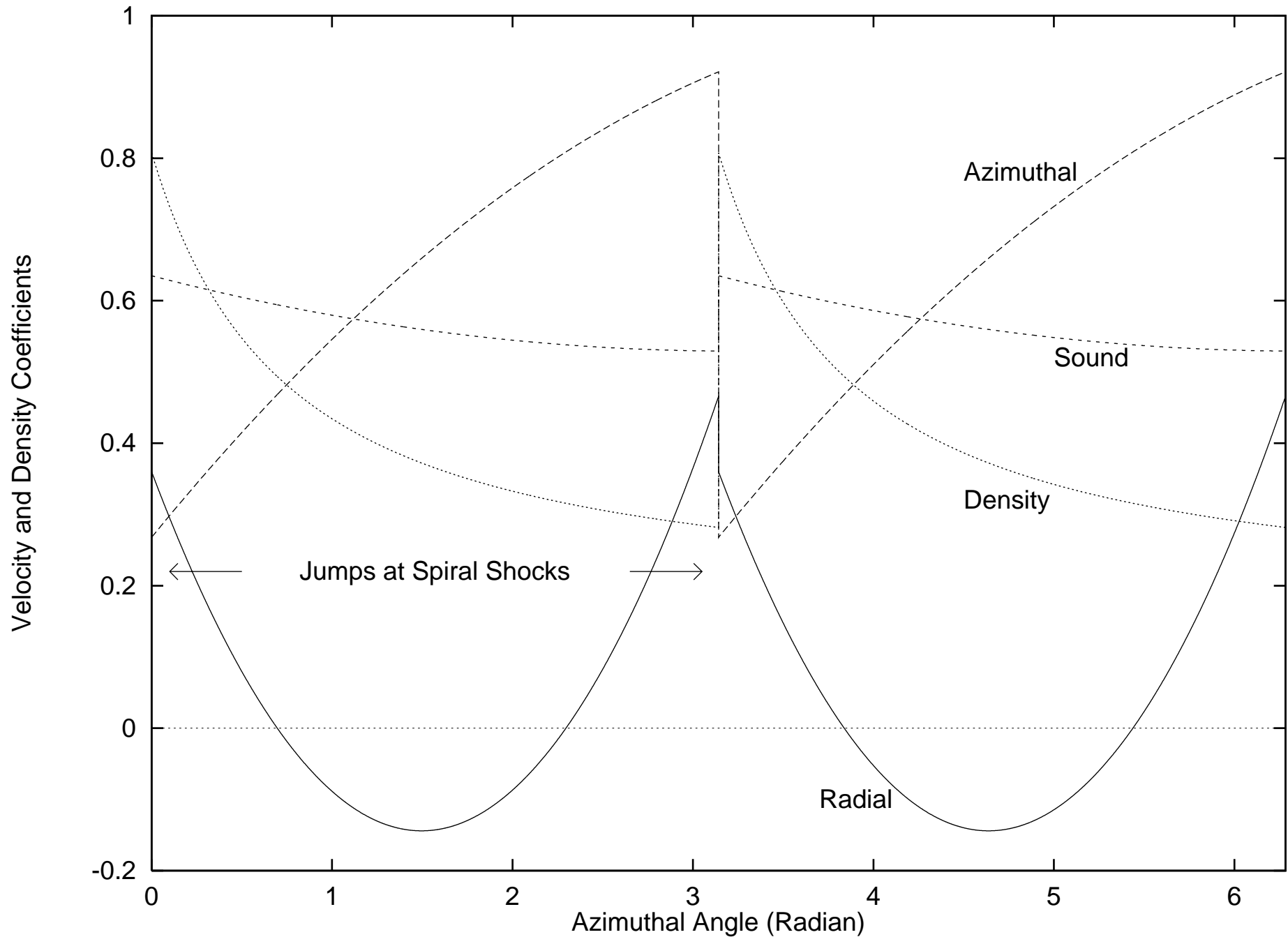
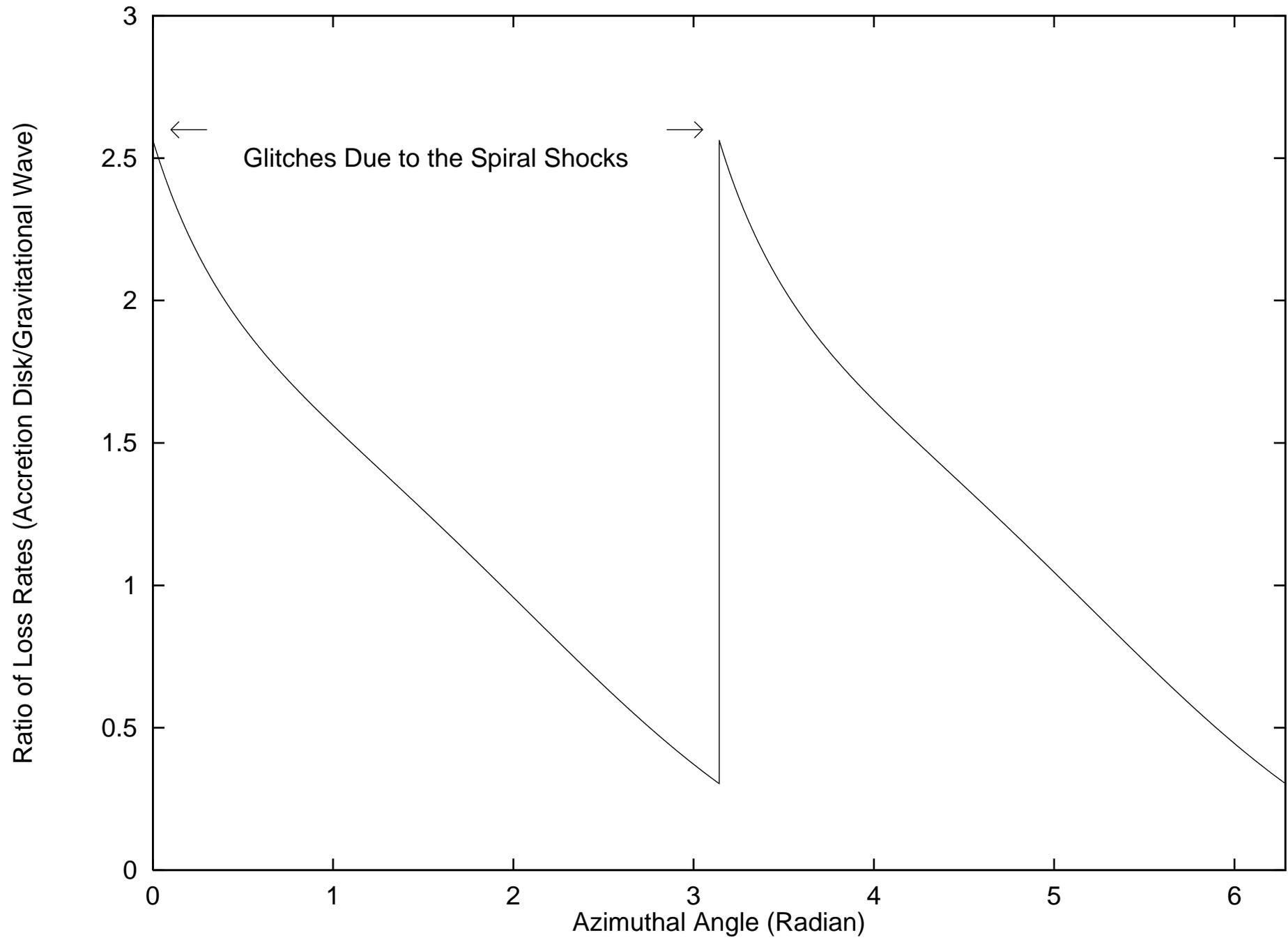
\end{document}